\documentclass[11pt]{article}
\title{ Skyrmions in the Quantum Hall effect and noncommutative solitons}
\author{V. Pasquier\\
CEA/Saclay, Service de Physique Th\'eorique\\
F-91191 Gif-sur-Yvette Cedex, FRANCE
}

\newcommand{\zbar}{\overline z}
\newcommand{\del}{\partial}

\begin{document}
\maketitle
\abstract{It has been recently shown that solitons 
are fundamental classical solutions of non-commutative field theories.
We reconsider this issue from the standpoint of the Hall effect and identify
some solutions with known solutions in the integer Hall effect with no Zeeman 
coupling. }

\section{Introduction}
\smallskip

This letter is made of two parts.

In the first part we show that the wave functions of two particles interacting
by a rotationally invariant potential in their Lowest Landau Levels is 
independent of the potential. This is well known \cite{LER}  when the two particles 
have equal and opposite charge and remains true when the charges are not equal.
The interest of this result is that the bound states formed by the two
particles are the relevant excitations of the fractional quantum Hall effect (FQHE).
In this case the two particles obey opposite
statistics and the bound states are the so called Composite fermions of the FQHE.
The universality of their
wave function is the seed for the rigidity of the many-body wave functions
of the FQHE \cite{JAIN} as explained in \cite{PAHA}. 
When the two charges are equal and opposite, the 
bound states form neutral
fermionic dipoles \cite{PAHA,READ}.
The field theory of these neutral dipoles is  related to the 
Fermi liquid observed at $\nu=1/2$ \cite{HALP}.
We briefly recall why the theory of these dipoles is expected to be noncommutative
(see also \cite{SUS,SUS1}).

In the second part
we study the condensation of bosonic dipoles made of two fermions
using the
techniques of \cite{PAHA}. 
These bosonic dipoles are useful to describe the quantum Hall effect
with spin
or the double layer systems.
This allows us to establish a link
between the skyrmion of the Hall effect 
\cite{REZ,SON}  and solitons
recently introduced in the context of non commutative string theory
\cite{GOP,MUK,HAR,WIT}.
The main physical difference 
lies in the extensiveness of the skyrmions
made out of a macroscopic number of dipoles whereas the noncommutative solitons
\cite{WIT}
coincide with their microscopic limit.

\section{Dipoles and displacement group}
\smallskip

In this section we consider the motion of two particles with
charges of opposite sign but not necessarily equal in magnitude
which interact attractively in a strong magnetic field.
If $V$ is a translation and rotation invariant
potential and $P_0$ projects
each particle in its Lowest Landau Level, 
the eigenstates of the matrix $P_0VP_0$ acting in the Hilbert space of the
two particles do not depend on $V$ and can be organized into
representations of a deformation of the displacement algebra.

Consider a charge $q>0$ particle in a magnetic field.
We use the symmetric gauge and take a units of length such
that the gauge field is given by: $A_x=y,\ A_y=-x$.
In terms of the  coordinates $z=x+iy, \ \overline z=x-iy$ 
one can define
two sets of mutually commuting oscillators:
\begin{eqnarray}
a=\del_{\zbar}+qz/2&,&\ a^+=-\del_{z} +q\zbar/2 \cr
b=\del_{z}+q\zbar/2&,&\ b^+=-\del_{\zbar} +qz/2 \cr
\nonumber
\end{eqnarray}

The first set of oscillators define the
free Hamiltonian $H=a^+a$ the  
eigenvalues of which correspond to the Landau levels (LL) $E_n=nq$ with $n\ge 0$. 
The
second set of oscillators $b,b^+$ are the guiding center coordinates. 
Together with angular the angular momentum $L=(b^+b-a^+a)/q$ 
they
generate a deformation of the
displacement algebra:
\begin{eqnarray}
\label{ALG}
[b,b^+]=q ,\
[L,b^+]=b^+ ,\
[L,b]=-b
\end{eqnarray}
and they commute with $H$ so that this algebra acts within each
LL. The representations are characterized by the eigenvalues $n \ge 0$ of
$b^+b/q-L$ which index the LL's and
the states $|n,l\rangle$ within a representation
are labeled by the angular momentum eigenvalues $l \ge -n$.
The wave functions in the Lowest LL $(n=0)$ are:
\begin{eqnarray}
\langle z|l\rangle=(q^{1/2}z)^l/(2\pi l!)^{1/2}e^{-qz\zbar/2}
\end{eqnarray}
It is helpful to visualize them as thin shells of radius $\sqrt{l/q}$
occupying an area $\pi/q$.  
The orbital degeneracy of a charge $q$ is thus $N_{\phi}=A/{2\pi l_q^2}$
where $A$ is the total area and $l_q=(2q)^{-1/2}$ defines the magnetic length.
It is also useful to introduce coherent states $|z\rangle$
defined by $\langle l|z\rangle=\langle z|l\rangle ^*$ which are the most localized states
in the Lowest Landau level (LLL).

Consider now two particles with positive charge $q_1,q_2$ interacting
by a translation and rotation invariant potential $V$ within 
their respective LLL ($n_1=n_2=0$).
If we label $1,2$ the oscillators associated to each particle, the operators
$B=b_1+b_2,\ B^+=b^+_1+b^+_2$ and $L=L_1+L_2$
act within the LLL and
commute with the interaction $V$.
They obey the relations (\ref{ALG}) with the parameter $q=q_1+q_2$.
Inside the LLL  the states $|n,l\rangle$ are labeled by
the eigenvalues of
$B^+B/q-L$ $n \le 0$   and the angular momentum 
eigenvalues $l \ge -n$. 
As a result the
eigenstates of $P_0VP_0$ where $P_0$ projects each particle in its
LLL do not depend on $V$ and their eigenvalues
$E_n$ are labeled by $n$. 

We are interested in the case where the first particle has a positive
charge $q_1$ and the second one has a negative charge $-q_2$, $q_1>q_2$.
In this case, $a^+_2$ and $b^+_2$ are an annihilation operator and it follows 
that the states associated to the second particle in the
$n^{\rm th}$ LL have their angular momentum $l_2\le n_2$.
When the two particles $1$ and $2$ interact
the operators $B,B^+,L$ obey the relations (\ref{ALG})
with the parameter $q^*=q_1-q_2$ and the states in the LLL $(n_1=n_2=0)$
$|n,l\rangle$ have $n\ge 0$ and $l \ge -n$.
They are the
eigenstates of $P_0VP_0$ and in the case where $V$ is attractive
they form
bound states which can be interpreted
as charge $q^*$ particles in the $ n^{\rm th}$ LL.
For example, the wave functions of the states with 
zero angular momentum ($l=0$) are given by:
\begin{eqnarray}
\phi_n(z_1,\zbar_2)=L_n(q^*z_1\zbar_2)
e^{q_2z_1\zbar_2-q_1z_1\zbar_1/2-q_2z_2\zbar_2/2}
\end{eqnarray}
where the $L_n$ are the Laguerre polynomials.
If the two charges are equal and opposite,  
$B$ and $B^+$ can be diagonalized simultaneously and play the role
of a generalized momentum.
The corresponding plane waves are then:
\begin{eqnarray}
\phi_{p,\bar p}(z_1,\zbar_2)=e^{q(z_1\zbar_2-
z_1\zbar_1/2-z_2\zbar_2/2)+i(\bar pz_1+p \zbar_2)}
\end{eqnarray}

This dynamical generation of higher Landau level wave functions or plane
waves inside the LLL plays
an important role in the FQHE. The bound states obtained in this way are called
composite 
particles and the field theory of these particles is non-commutative as we
show now (see also \cite{PAHA,READ}).
Let us consider the case where the two charges are equal in magnitude $q_1=q_2$.
Then the bound states form neutral dipoles and 
the non-commutativity originates from their extensiveness \cite{PAHA,READ,SUS1}.
The projection into the LLL
transforms $f(\vec x)$ into a matrix : 
$$ \hat f_{l,l'}=
\langle l|f| l' \rangle$$
where the matrix elements are taken in between LLL orbitals.
The action of this matrix on the state $|l\rangle$ being given by
$\sum_{l'} f_{l,l'}|l'\rangle$.
The function $f(z,\zbar)$ is called the P-symbol for the matrix  $\hat f$
\cite{KLAU}. If we use the coherent state basis to represent the LLL orbitals one can rewrite the
matrix $\hat f$ as: 
$$\hat f=\int\ |z\rangle f_p(z,\zbar)\langle z|\ dz d\zbar.$$
and because the LLL basis is not complete,
the matrices associated to different functions not commute \cite{PAHA,READ}. 

Another way to associate a function to a matrix consists in
bracketing the matrix between coherent states:
\begin{eqnarray}
f_q(z,\zbar)&=&\langle z| \hat f|z\rangle 
\label{COMU}
\end{eqnarray}
It is called the Q-symbol and will be
useful when we 
consider fields in the next section. 
In particular one has:
\begin{eqnarray}
e^{i(\bar p z+p\zbar)}
=\langle z|e^{i{\bar p b^+\over q}} e^{i {p b\over q}}|z\rangle 
\label{COMU1}
\end{eqnarray}
The Q-symbol induces a noncommutative product on functions which we denote by $*$:
$f_q *g_q=(\hat f \hat g)_q$. In the following, the Q-symbol is understood
whenever we consider a *-product.

In the dipole case
there are $4$ type of operators to consider according to
how they intertwine the particles $1$ and $2$.
They can be represented in the matrix form:
\begin{eqnarray}
\pmatrix{\hat f_{11}&\hat f_{12}\cr
\hat f_{21}&\hat f_{22}\cr}
\label{TEST}
\end{eqnarray}
where each matrix element is a $N_{\phi}^2$ matrix. We can also consider the Q-symbol
which is a two by two matrix with functions of $z,\zbar$ as matrix elements.

\section{Skyrmions and non-commutative solitons} 

The simplest model for dipoles occurs  for a system of $N_e$
electrons interacting repulsively
at $\nu=1$: $N_e=N_{\phi}$. The electrons carry a spin $1/2$
which is not coupled to the magnetic field in the limit of zero
Zeeman coupling. At this filling the ground state is a ferromagnet
with total spin $N_e/2$ where $N_e=N_{\phi}$ is the total number of electrons
in the system. 

The addition or removal of a single
electron dramatically reduces the spin as seen numerically \cite{REZ}
and experimentally \cite{TYC}.
This behavior may be explained using the so called skyrmion spin textures
\cite{SON,MAC} in
a non linear $\sigma$ model approach.
These textures carry a topological number equal to the electric charge.
In this section 
we review these solutions and reinterpret them as
non-commutative solitons.

Let us see heuristically how this problem is related to the dipoles and
why we expect them to
condense in presence of an electric charge.
The picture we have in mind is a ferromagnetic ground state where all spins
are aligned in the minus direction. Each time we flip a spin a spin one
dipole is formed made by the spin up electron and the hole created in the 
spin down electron sea.
If we remove an electron from the  ground state, a condensate will
form if it is energetically favorable to flip a large number of spins to
lower the energy. In the dipole language it means that a large number of dipoles 
agglomerate around the charge. To see that this occurs
consider one dipole in presence of a positive charge equal in magnitude to the
charges of the dipole. It is equivalent to consider
two positive charges denoted $1,2$ in presence of a negative 
charge denoted $3$. The charges with the same sign repel each other through the
interaction $V$ while the charges with opposite sign attract each other through $-V$. 
In the strong field limit we can neglect the masses of the particle
and by the canonical quantization procedure the coordinate $R_{13}=R$
becomes conjugated to $R_{23}=P^{\perp}$, $[P^{\perp}_x,R_y]=-[P^{\perp}_y,R_x]=1$.
The Hamiltonian which describes the three charges is the sum of their interactions:
$H=V(R-P^{\perp})-V(R)-V(P^{\perp})$. For a reasonable repulsive potential a
bound state forms with the three particles aligned and the negative charge
in the middle of two positive charges (see \cite{DZY} for a more complete treatment). 
The angular momentum is proportional
to the area since $L=R.P^{\perp}$.
Since a single dipole binds to the charge
we expect that 
several (bosonic) dipoles condense in the presence of the charge
and that the condensate carries an angular momentum $L$ proportional to its mean square
radius.

To make the discussion simplest we consider the case 
of a hard core potential $V=\delta^{(2)}(\vec x)$ and
the electrons live on a sphere of area $(2L+1)\pi$
thread by $2L$ quantum fluxes where $L$ is half integer.
The LLL is $2L+1$ degenerate and in the azimuthal gauge $a_{\phi}=1-\cos(\theta)$ 
the LLL orbitals take the form:
\begin{eqnarray}
\langle \hat r|l\rangle={2L \choose l+L}^{1/2} u^{L+l} v^{L-l} 
\label{FORM}
\end{eqnarray}
$-L \le l \le L$ where $u=\sin(\theta/2)e^{i\phi}$ and 
$v=\cos(\theta/2)$.

The zero energy 
wave functions are
homogeneous polynomial of
degree $2L$ in the coordinates $u_i,v_i$
of the electrons
and vanish when two particles are at the same position.
When there are exactly $2L+1$ electrons on the sphere
the Pauli principle forces it to be symmetric under the spin permutations
and the ground state is thus ferromagnetic.
If we remove $1$ electron from the system, Fertig et al. \cite{MAC} proposed
a wave function satisfying these constraints :
\begin{eqnarray}
\Phi=\prod_{i=1}^{N_e}(u_i |\downarrow \rangle + 
v_i |\uparrow\rangle )\prod_{i<j} (u_i v_j- u_j v_i)
\end{eqnarray}
The effect of the first factor is to expel the spin down electrons from the first orbital $(l=-L)$
and the spin up electron from the last orbital $(l=L)$
so that the spin is up at the north pole and down at the south pole.
This wave function being invariant under simultaneous space and spin rotation, it follows
that the spin takes a hedgehog shape around the sphere.
Although it is not a spin eigenstate, it has its maximal weight for the spin and
angular momentum $S=L=0$.
It fits our qualitative picture where on a sphere of area $(2L+1)\pi$,
$L$ dipoles bind 
to the charge to form a bound state of spin and the angular 
momentum $L$.



To relate this discussion to the noncommutative field theory it is useful to introduce
the fermionic operators which create a spin up or down
electron in the $l^{\rm th}$ orbital:
$c^+_{l\uparrow} ,c^+_{l\downarrow} $.
The
products  $c^+_{l\uparrow}c_{l'\downarrow}$  create a 
spin one dipole made of charge $q=1$ particle in the $l$ orbital
and a charge $q=-1$ hole in the $l'$ orbital.
It is possible to express the 
relevant observables
in terms of a fundamental field which creates the dipoles.
For this we define the matrix fields
$\sigma_{ij},\bar \sigma_{kl}$ which obey canonical commutation relations:
\begin{eqnarray} 
&&[\sigma_{ij},\sigma_{kl}]=
[\bar \sigma_{ij},\bar \sigma_{kl}]=0
\nonumber \\
&&[ \sigma_{ij},\bar \sigma_{kl}]=\delta_{il} \delta_{jk}
\label{FERMCO}
\end{eqnarray}
Next we identify fields
in the two representations by requiring that they obey the same commutation relations:
\begin{eqnarray}
A= 
\pmatrix{c^+_{i\downarrow}c_{j\downarrow}&c^+_{i\uparrow}c_{l\downarrow}\cr
        c^+_{k\downarrow}c_{j\uparrow}&c^+_{k\uparrow}c_{l\uparrow}\cr}=
\pmatrix{(\bar \sigma\sigma)_{ij}&(\bar \sigma.\sqrt{1-\sigma\bar \sigma})_{il}\cr
        (\sqrt{1-\sigma\bar \sigma}.\sigma)_{kj}&(1-\sigma\bar \sigma)_{kl}\cr}
\label{RHO}
\end{eqnarray} 
Indeed, the matrix elements of the left-hand side matrix obey the commutation relations of
a $U(2N)$  Lie algebra and the right-hand side matrix can be viewed as a generalized
Holstein-Primakov representation for these generators \cite{PAHA}.
If we go to the Q-representation, the equality (\ref{RHO}) translates 
into an identification between fields:
\begin{eqnarray}
\pmatrix{\phi^+_{\downarrow}\phi_{\downarrow}&\phi^+_{\uparrow}\phi_{\downarrow}\cr
\phi^+_{\downarrow}\phi_{\uparrow}&\phi^+_{\uparrow}\phi_{\uparrow}\cr}(\hat r)
=
\pmatrix{\bar \sigma*\sigma&\bar \sigma*\sqrt{1-\sigma*\bar \sigma}\cr
        \sqrt{1-\sigma*\bar \sigma}*\sigma&1-\sigma*\bar \sigma\cr}(\hat r)
\label{HOL}
\end{eqnarray} 
where the fields  $\phi^+_{\uparrow,\downarrow}(\hat r)=\sum_l  \langle \hat r|l\rangle
c^+_{l\uparrow,\downarrow}$  create an electron up or down at a given position.
The diagonal matrix elements are the density of up and down electrons.
In the case of a $\delta$ potential 
the energy is
the integral of the total density squared
which in term of $\sigma(\hat r)$ is given by:
\begin{eqnarray}
E=\int (\sigma*\bar \sigma-\bar\sigma*\sigma -1)^2  d^2 \hat r
\end{eqnarray}
We then minimize the energy for a classical field 
and we thus replace $\sigma_{ij}$ by a c-number in this expression.
The minimum is clearly for a  constant density:
\begin{eqnarray}
[\sigma,\bar \sigma]={1\over 2L+1}
\end{eqnarray}
A solution valid in the north hemisphere is $\sigma_{l-1,l}=\sqrt{l+L\over 2L+1}$.
The expression which follows for $A$ is then:
\begin{eqnarray}
A={1\over 2L+1}
\pmatrix{L+l&L_+\cr 
L_-&L-l\cr}
\label{MATR}
\end{eqnarray}
where $\vec L$ is the angular momentum operator.
In the south hemisphere we must use the gauge $\tilde a_{\phi}=1+\cos(\theta)$.
The LLL basis $|\tilde l\rangle$ takes the form (\ref{FORM})
where $\tilde u=\cos(\theta/2)e^{-i\phi}$ and 
$\tilde v=\sin(\theta/2)$. In the intermediate region the two basis are related
by a gauge transformation 
$e^{2iL\phi}\langle \hat r|\tilde l\rangle=\langle \hat r|l\rangle$ with $l=-\tilde l$.
In the basis $|\tilde l\rangle$ the expression (\ref{MATR}) of $A$
must be conjugated by $\sigma_x$.
Thus we use the same expression 
$\tilde \sigma_{ll'}=\sigma_{ll'}$ in the right hand side of (\ref{RHO})
to represent $\sigma_x A \sigma_x$ in the south hemisphere.
Clearly we cannot use a single matrix $\sigma$
to represent $A$ globally since the trace of 
the left-hand side of (\ref{RHO})
must be equal to the number of electrons $2L$ and the trace of the right-hand side
is equal to $2L+1$ for any finite size matrix $\sigma$.

The Q-symbol expression of $A$ at scales large compared to the magnetic length 
is obtained by substituting in (\ref{HOL}) $\sigma_q (\hat r)
\simeq \sin(\theta /2)e^{i\phi}$.
It has the characteristic hedgehog shape
of the spin texture. 

Recently, non-commutative solitons were introduced in relation with strings \cite{GOP}.
The solution presented here has many similarities with the one presented in \cite{WIT}
following \cite{MUK,HAR}.
In particular the field $A$ is a projector $A^2=A$.
There is however a major physical difference: 
on the plane, the equivalent of the skyrmion \cite{MAC} has a parameter
which controls its size.
The solution $A$ of \cite{WIT} coincides with the 'zero size soliton' as can easily be seen
by interpreting its Q-symbol as a spin texture.
In the case of the $\delta$ potential, the energy of the skyrmions
is independent of their size
but as soon as some nonlocal repulsion is introduced large skyrmions are 
energetically favored. 

One of the aim of this letter was to establish a link between an
experimentally observed phenomena in the integer Hall effect
and classical solutions of 
noncommutative field theories.
In the context of the fractional Hall effect
the dipoles that have been considered in \cite{PAHA,READ} are fermions and
the field $\sigma$ is thus fermionic. 
In this case  nonlinear classical solutions, if they exist,
may be related
to the striped phases \cite{KOU} (see \cite{SO2} for a proposal
in this direction) or pair condensation \cite{MOO}
predicted for higher Landau levels.

\smallskip
I thank D.Bernard, J.Chalker, G.Misguich, U.Moschella and D.Serban for
discussions and advices.

\vfill\eject
\end{document}